\documentclass[pdflatex,sn-mathphys-num,twocolumn]{sn-jnl}

\usepackage{graphicx}%
\usepackage{booktabs} 
\usepackage{amsmath,amssymb,amsfonts}%
\usepackage{amsthm}%
\usepackage{mathrsfs}%
\usepackage[title]{appendix}%
\usepackage{xcolor}%
\usepackage{textcomp}%
\usepackage{manyfoot}%
\usepackage{booktabs}%
\usepackage{algorithm}%
\usepackage{algorithmicx}%
\usepackage{algpseudocode}%
\usepackage{listings}%
\usepackage{multirow} 

\theoremstyle{thmstyleone}%
\theoremstyle{thmstyletwo}%

\theoremstyle{thmstylethree}%

\begin{document}

\title[Article Title]{Tree-NET: Enhancing 2D medical image segmentation through efficient low-level feature training}


\author*[1,2]{\fnm{Orhan} \sur{Demirci}}\email{orhandemirci@cs.hacettepe.edu.tr}

\author[3]{\fnm{Bulent} \sur{Yilmaz}}\email{yilmaz.b@gust.edu.kw}

\affil*[1]{\orgdiv{Electrical and Computer Engineering Department}, \orgname{Abdullah Gul University}, \orgaddress{ \city{Kayseri}, \postcode{38080},  \country{Turkey}}}

\affil[2]{\orgdiv{Computer Engineering Department}, \orgname{Hacettepe University}, \orgaddress{\street{}, \city{Ankara}, \postcode{06800}, \country{Turkey}}}

\affil[3]{\orgdiv{Electrical and Computer Engineering Department}, \orgname{Gulf University for Science and Technology (GUST)}, \orgaddress{ \city{Hawally}, \postcode{32093}, \country{Kuwait}}}

\abstract{This paper introduces Tree-NET, a novel framework for medical image
segmentation that leverages bottleneck supervision to enhance both
segmentation accuracy and computational efficiency. While previous studies have applied bottleneck feature supervision to segmentation tasks, it has typically been limited to the training phase, offering no computational benefits during inference. To the best of our knowledge, this is the first framework to employ dual bottleneck supervision for segmentation, leveraging latent space features at both the input and output stages. This approach reduces input and label dimensions with minimal parameter overhead, while preserving accuracy. Tree-NET features a three-component architecture: Encoder-Net and Decoder-Net, which compress input and label data via autoencoding, and Bridge-Net, a segmentation model trained on these compressed representations. By operating entirely on dense, low-dimensional features, Tree-NET improves runtime efficiency and can be integrated into existing segmentation models without modifying their internal structures or increasing model size. We evaluate Tree-NET on two
key segmentation tasks—skin lesion and polyp segmentation—using various
backbone models, including U-NET, U-NET++, and Polyp-PVT. Experimental results show that Tree-NET reduces FLOPs by a factor of 4 to 13 and decreases memory usage, while maintaining segmentation accuracy comparable to baseline models. For example, with an untrained U-NET++ backbone, Tree-NET improves the Dice score on ISIC-2018 from 0.829 to 0.862 and the IoU from 0.736 to 0.787. On CVC-ClinicDB, it achieves a Dice of 0.946 and an IoU of 0.901 using a Polyp-PVT backbone, matching or surpassing baseline performance. These findings underscore Tree-NET’s potential as a robust and efficient solution for medical image segmentation. 
}




\maketitle

\section{Introduction}\label{sec1}
Segmentation of medical images presents significant challenges compared to natural image segmentation due to irregularities in the size, shape, and intensity of target objects, high rates of false positives, and the cost and time required for annotation \cite{zhao2020rethinking}. These issues are further compounded by the scarcity of annotated datasets and the stringent clinical precision required, as even minor segmentation errors can adversely impact diagnosis and treatment outcomes.

Traditional segmentation models, such as U-NET and variants of U-NET \cite{ronneberger2015u,zhou2018unet++}  and fully convolutional networks (FCNs) \cite{ji2020parallel}, have been the backbone of medical image segmentation. These architectures, built on encoder-decoder designs, utilize skip connections to combine coarse-grained semantic features with fine-grained spatial details, enabling accurate delineation of complex structures. 




More recently, transformer-based architectures, such as Pyramid Vision Transformers (PVT) \cite{wang2021pyramid} , Polyp-PVT \cite{dong2021polyp}, TransUNet \cite{chen2021transunet, chen2024transunet} and Swin UNet \cite{cao2022swin}, have emerged as powerful alternatives in medical image segmentation. Unlike convolutional models, transformers leverage attention mechanisms to capture global contextual information, enabling superior performance on tasks requiring both fine-grained and long-range feature representation. For instance, TransUNet incorporates transformers into its encoder, capturing long-range dependencies while retaining critical spatial details via skip connections. Similarly, Swin UNet employs hierarchical attention mechanisms, improving computational efficiency and performance across diverse datasets. Polyp-PVT, designed specifically for polyp segmentation, excels in delineating irregular and complex boundaries.

Despite their promise, transformer-based models face challenges such as heavy computational demands, substantial memory requirements, and limited interpretability, all of which hinder their clinical applicability \cite{butoi2023universeg}. These limitations underscore the pressing need for segmentation models that effectively balance accuracy, efficiency, and generalizability.

One major challenge is the significant computational burden, especially in 3-D tasks like volumetric brain or lung segmentation \cite{xiao2023transformers}, which leads to increased processing times and substantial demands on hardware resources, making real-time analysis difficult. High-resolution medical images and complex architectures also result in excessive memory consumption, often causing out-of-memory errors. Additionally, achieving high accuracy with transformer-based models typically requires extensive datasets and prolonged training periods, which are compounded by the scarcity and cost of annotated medical images. Transformers' reliance on large datasets for effective training poses a major limitation in the field, where data is often limited. Furthermore, their high parameter counts increase the risk of overfitting, particularly on small datasets, a common scenario in medical imaging.

Generalization is another concern, as models trained on specific datasets often perform poorly when applied to different medical imaging contexts due to variations in imaging techniques, patient populations, and disease presentations \cite{butoi2023universeg}. Biases present in training data can further reduce their reliability across diverse clinical applications. Interpretability remains a critical issue, as understanding the decisions of transformer models is essential for clinical validation and trust. Lastly, their integration into clinical workflows is challenging, as their outputs frequently require adaptation to meet specific clinical requirements.

\begin{figure*}[h]
	\centering
	\includegraphics[width=1.8\columnwidth]
 {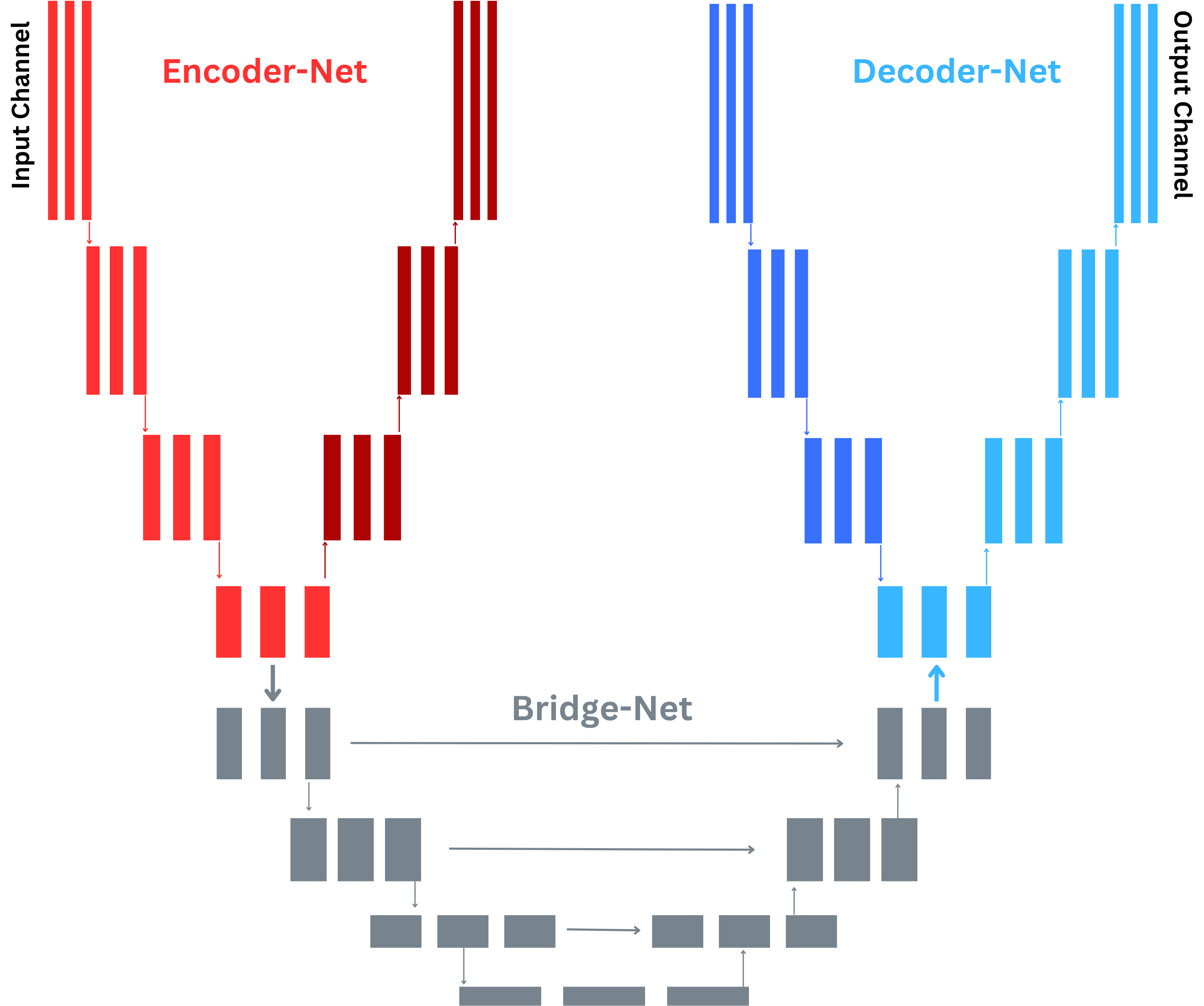}
	\caption{General Tree-NET diagram. Red blocks denote the Encoder-Net for input encoding. Gray blocks represent the U-NET architecture as the Bridge-Net for segmentation training. Blue blocks indicate the Decoder-Net for output decoding to match the original label size. Darker shading within the Encoder-Net and Decoder-Net indicates components that are removed after training and are not included in the final assembled inference network.}
	\label{FIG:1}
\end{figure*}

To address these issues, Tree-NET, a novel state-of-the-art architecture, is proposed for medical-image segmentation tasks based on bottleneck feature supervision of both input and label masks and then creating a bridge network between them to train their low-level representations (Figure \protect\ref{FIG:1}). The network consists of three main components named Encoder-Net, Decoder-Net and Bridge-Net as each having separate training processes. After the trainings of each component are completed, all three components are combined end-to-end and available for the segmentation task. By this technique, the bridge network of choice is trained with smaller sized input and label images without losing any essential features improving the model performance in terms of accuracy and computational cost while keeping the trainable parameters almost the same number.

In this study, we introduce Tree-NET, a novel deep learning architecture with a unique autoencoder-supported design that significantly enhances segmentation performance while reducing computational costs. Our key contributions are as follows:

\begin{itemize} \item We propose Tree-NET, a deep learning model that employs bottleneck feature supervision to refine segmentation training. This architecture leverages the low-level representations of input and label data to improve segmentation outcomes.

\item Tree-NET features an end-to-end structure composed of three distinct components—Encoder-Net, Bridge-Net, and Decoder-Net. Each component undergoes isolated training before integration, optimizing feature learning and model performance.

\item The model utilizes Encoder-Net and Decoder-Net to supervise input and label data, respectively, before the segmentation training conducted by Bridge-Net. This approach enhances feature learning by allowing for more learning steps and a more coordinated architecture. Additionally, it provides the flexibility to use different hyperparameters for each component, optimizing performance.

\item Bridge-Net, a key component of Tree-NET, can be implemented using various existing segmentation algorithms. This flexibility allows Tree-NET to adapt to different segmentation methods and requirements, making it a versatile tool for various applications.

\item Tree-NET reduces computational costs by shrinking input and label sizes without loss of critical information. This results in decreased memory usage and improved processing speed, all while enhancing segmentation accuracy. \end{itemize}

The remainder of this paper is structured as follows: Section 2 reviews related work on bottleneck feature supervision and transformer-based segmentation architectures. Section 3 describes the Tree-NET architecture, detailing the Encoder-Net, Bridge-Net, and Decoder-Net components. Experimental settings, datasets, and evaluation metrics are discussed in Section 4, followed by a presentation of results in Section 5. Section 6 offers a discussion of efficiency advantages, implementation challenges, and potential improvements. Finally, Section 7 provides conclusions and future research directions.

\section{Related work}\label{secc2}

Deep encoder–decoder CNNs such as U-NET \cite{ronneberger2015u} and U-NET++ \cite{zhou2018unet++} have become the de-facto standard for medical image segmentation, but their high memory footprint and FLOP count limit deployment in real-time or resource-constrained settings.  To alleviate this computational burden, recent research has focused on \textit{bottleneck supervision}: guiding or redesigning the low-dimensional latent representations inside a network so that essential information is retained while redundant computation is removed.

The literature falls into two main groups.  
(i)~\textbf{Input-side supervision} compresses the \emph{input} feature maps—often with an auto-encoder—so that the downstream network operates entirely on a compact latent space.  
(ii)~\textbf{Label-side supervision} compresses the \emph{ground-truth labels} and uses those compressed targets to guide the segmentation network’s bottleneck.  
Tree-NET combines both ideas, supervising bottlenecks at \emph{both} ends and thus achieving dual compression.

\subsection{Input-side bottleneck supervision}
Recent work in hyperspectral \emph{classification} shows that unsupervised
auto-encoder architectures can compress high-dimensional inputs, cutting
computational cost while retaining discriminative information.  Although these
methods are not yet applied to segmentation, they illustrate how strong task performance can be maintained on severely reduced inputs.

Mei \emph{et al.}\ \cite{mei2019unsupervised} proposed a 3-D convolutional
auto-encoder (3D-CAE) that captures spatial–spectral correlations and
markedly reduces dimensionality without notable information loss.  
Zhao \emph{et al.}\ \cite{sun2021unsupervised} extended this idea with a
multi-level 3D-CAE, extracting features from several encoder depths to preserve
multi-scale context while remaining efficient.  
Bai \emph{et al.}\ \cite{bai2024two} introduced a two-stage
multidimensional stacked auto-encoder (TMC-SAE): a 1-D spectral compressor
followed by 2-D/3-D convolutions for spatial–spectral fusion, achieving higher
accuracy at lower compute.  
Most recently, Lin \emph{et al.}\ presented \emph{TRACE}\,\cite{orouji2025task},
which learns a low-dimensional input bottleneck supervised by a logistic-regression
head; labels remain uncompressed and no decoder is used, further confirming that
carefully guided input compression can lower FLOPs without harming task
performance.

These studies demonstrate that input-side bottleneck supervision is a viable
route to efficiency—an idea that Tree-NET extends from classification to
segmentation by pairing an input compressor with a complementary
label-side bottleneck.

\subsection{Label-side bottleneck supervision}

A complementary line of research supervises the network’s bottleneck
representations using information derived from ground-truth labels.  This
strategy—often referred to as a Teacher–Learner (T–L) or Target–Label
framework—typically unfolds in two stages.  First, an auto-encoder (the
\emph{Teacher}) is trained solely on label masks to produce compact target
vectors (bottleneck features) \cite{oktay2017anatomically}.  Second, a primary
segmentation network (the \emph{Learner}) is trained on input images with an
auxiliary loss that encourages its own bottleneck representations to match
those precomputed targets.

A notable example is the Bottleneck Feature-Supervised U-NET (BS U-NET)
proposed by Song \emph{et al.} \cite{song2020bottleneck}.  BS U-NET employs an
\emph{Encoding U-NET}, trained only on labels, to generate the target
bottleneck features that supervise a conventional Segmentation U-NET.  The
overall loss combines the standard segmentation objective with a bottleneck
matching term.

However, the Encoding U-NET is discarded after training, and only the standard
Segmentation U-NET is used for inference.  As a result, BS U-NET provides no
runtime gains in FLOPs or memory usage; the deployed architecture is identical
to the original U-NET, and accuracy improvements remain modest.

\subsection{Tree-NET’s approach.}  
Tree-NET builds upon these concepts by integrating dual bottleneck supervision—compressing both inputs and labels—within a three-stage architecture tailored for segmentation. To our knowledge, Tree-NET is the first framework to apply such a dual compression strategy in a segmentation context, operating entirely within compressed feature spaces to achieve significant reductions in computational requirements without compromising accuracy. Our comprehensive literature review, including recent works from 2023–2025, did not identify any prior models employing both input- and label-side bottleneck supervision together in a unified, three-stage pipeline.

Our proposed Tree-NET framework builds upon the concept of bottleneck supervision but introduces a fundamentally different architecture aimed at enhancing both computational efficiency and segmentation accuracy. Tree-NET employs two dedicated autoencoders—Encoder-Net for inputs and Decoder-Net for labels—each trained independently to compress input images and ground-truth masks into low-dimensional bottleneck representations. The central segmentation network, Bridge-Net, operates entirely on these compact feature spaces and is supervised using the bottleneck outputs of the Decoder-Net. Importantly, the encoder of Encoder-Net and the decoder of Decoder-Net are retained during inference and integrated with Bridge-Net to reconstruct full-resolution segmentation maps. This distinct three-stage pipeline (Encoder-Net → Bridge-Net → Decoder-Net) significantly reduces FLOPs and memory usage during both training and inference, while maintaining—or even improving—segmentation accuracy compared to conventional full-resolution architectures.

\section{Model architecture}\label{sec2}

The Tree-NET model consists of three main components such as Encoder-Net, Bridge-Net, and Decoder-Net.
\subsection{Encoder-Net}
\begin{figure}[h] 
    \centering
    \includegraphics[width=0.9\columnwidth]{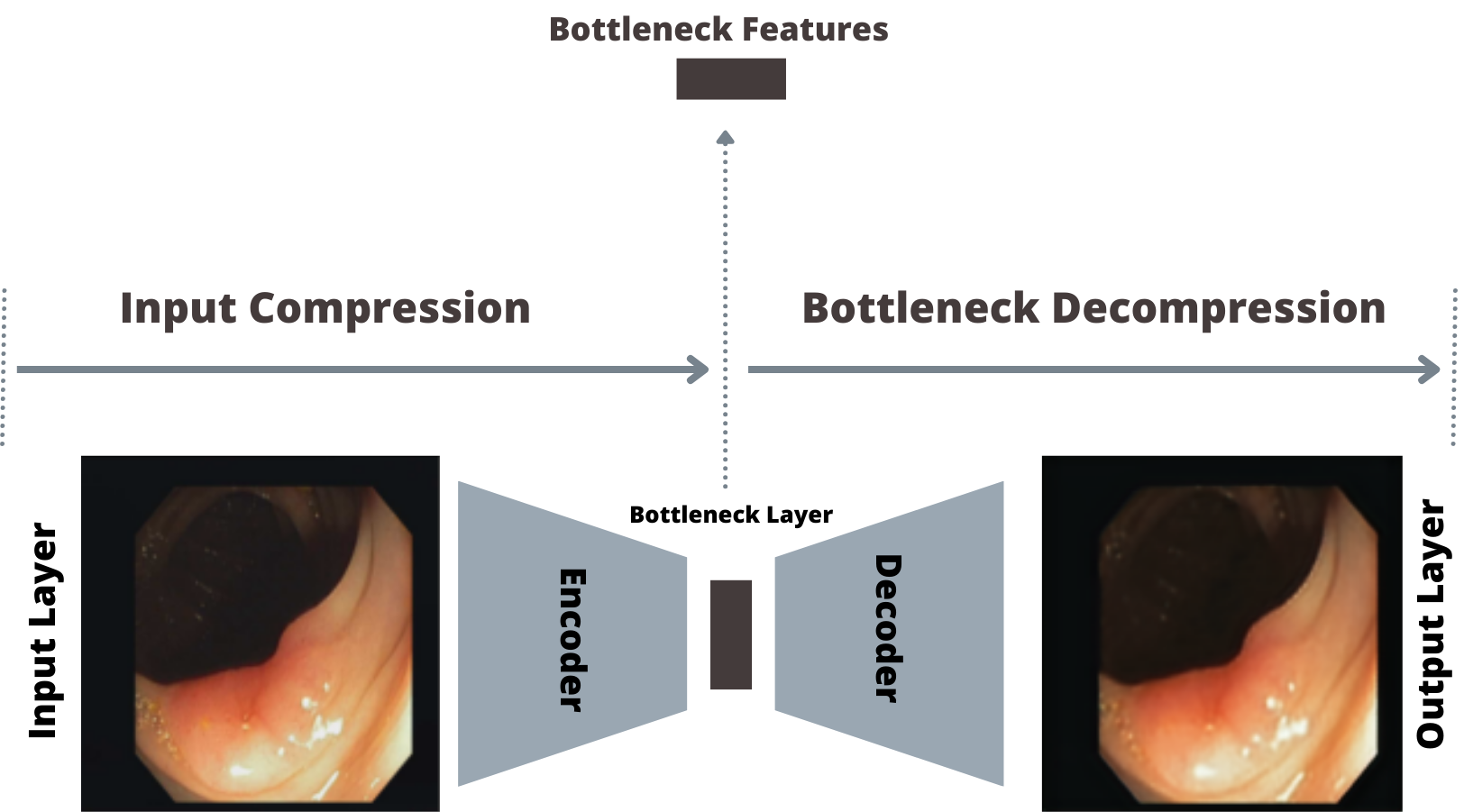} 
    \caption{Schematic representation of Encoder-Net.}
    \label{FIG:Encoder}
\end{figure}

Encoder-Net is a convolutional autoencoder model consisting of encoder and decoder parts (see Figure \protect\ref{FIG:Encoder}).

It is designed to extract low-level representations (bottleneck features) from input images. During training, the Encoder-Net processes the input images to generate these bottleneck features, which are then fed into the Bridge-Net for segmentation training. This means that, once training is complete, only the encoder part of the Encoder-Net is used. (See Figure \protect\ref{FIG:Tree-NET Diagram})).

\subsection{Bridge-Net}
\begin{figure}[h] 
    \centering
    \includegraphics[width=0.9\columnwidth]{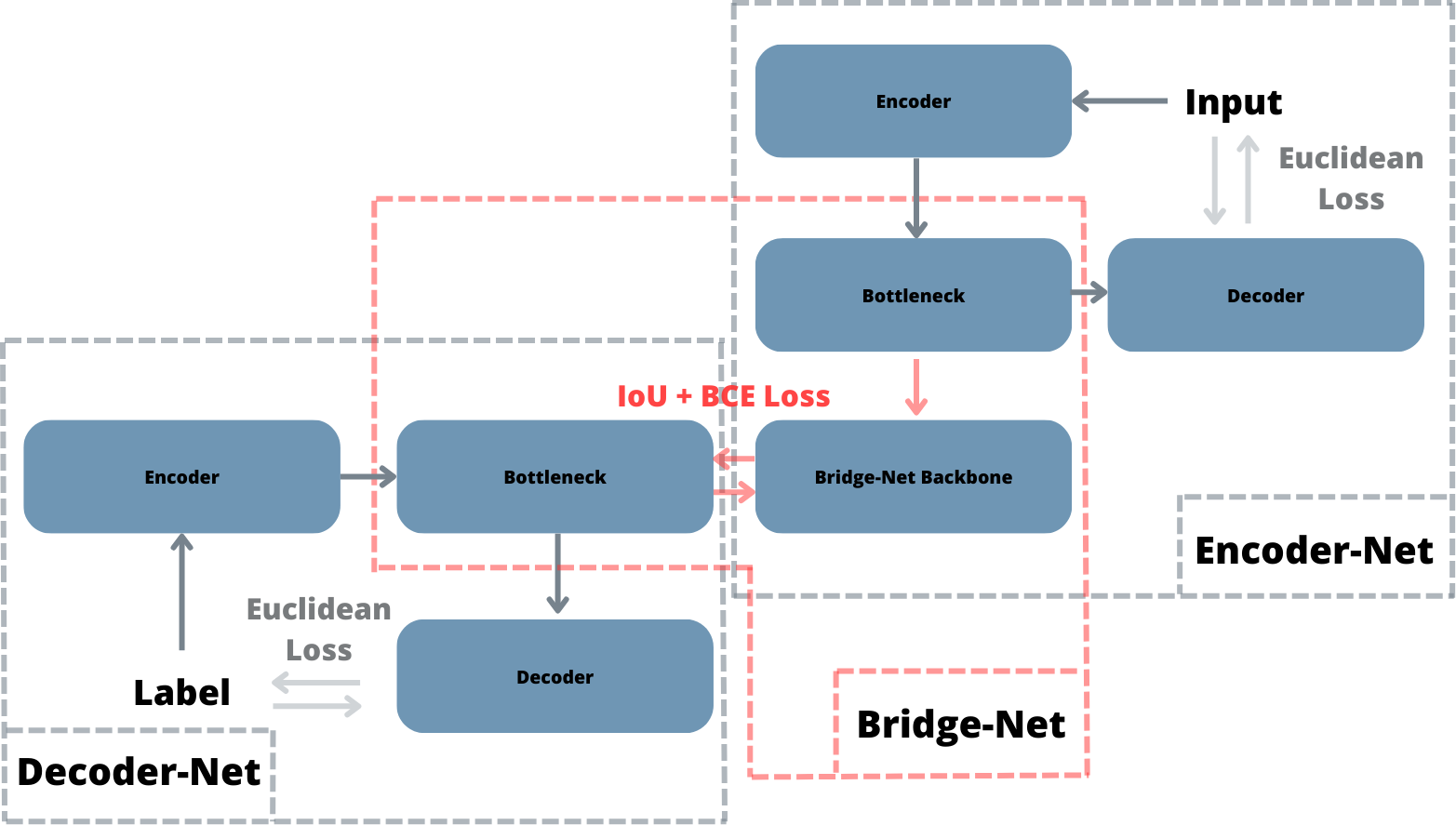} 
    \caption{Tree-NET Components: The diagram illustrates the Encoder-Net, Bridge-Net, and Decoder-Net. Unidirectional arrows indicate the data flow through the networks, while bidirectional arrows represent the process of loss computation.}
    \label{FIG:Bridge}
\end{figure}

As the name implies, Bridge-Net is the network where the Encoder-Net and Decoder-Net are integrated. In this model, the bottleneck features from the Encoder-Net are fed into the Bridge-Net while the bottleneck features of Decoder-Net are used as the labels. Figure \protect\ref{FIG:Bridge} explains the process. Bridge-Net is a framework that any type of segmentation model can be used for this purpose. In our experiment, we use U-Net, U-Net++ and Polyp-PVT as the Backbone (BB) models for the Bridge-Net in segmentation tasks.

\subsection{Decoder-Net}
\begin{figure}[h] 
    \centering
    \includegraphics[width=0.9\columnwidth]{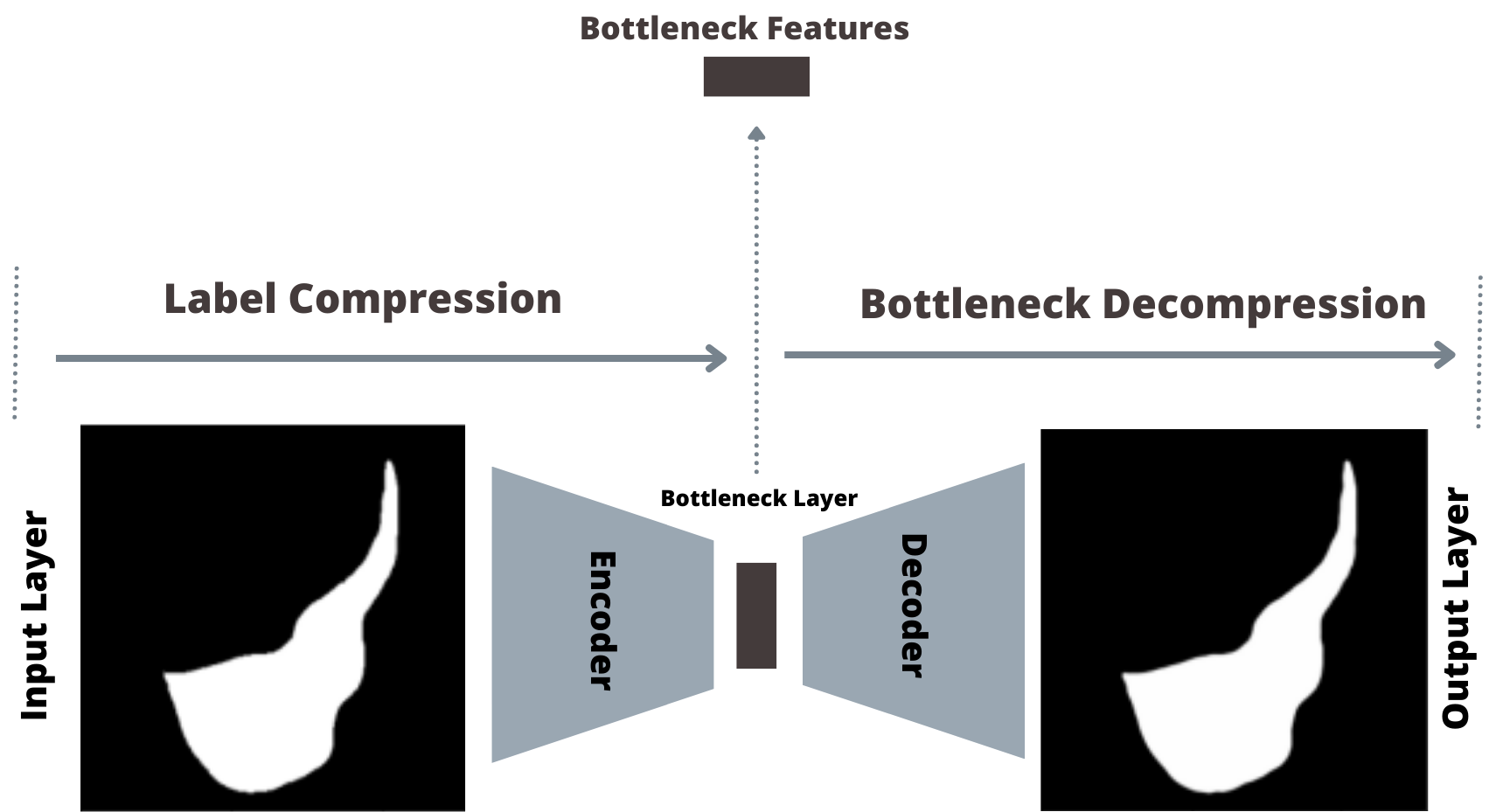} 
    \caption{Schematic representation of Decoder-Net.}
    \label{FIG:Decoder}
\end{figure}

Decoder-Net has a similar structure with the Encoder-Net. The main difference is that it extracts the low-level representation (bottleneck feature) of labels masks. The Decoder-Net training is separately applied on label images and bottleneck features obtained for training the Bridge-Net (Figure \protect\ref{FIG:Decoder}). 
After training is complete, only the decoder part of the Decoder-Net is used to convert the output back to its original size, producing the final segmentation result. (See Figure \protect\ref{FIG:Tree-NET Diagram}).

\subsection{Overall architecture}
\begin{figure*}[h] 
    \centering
    \includegraphics[width=1.8\columnwidth]{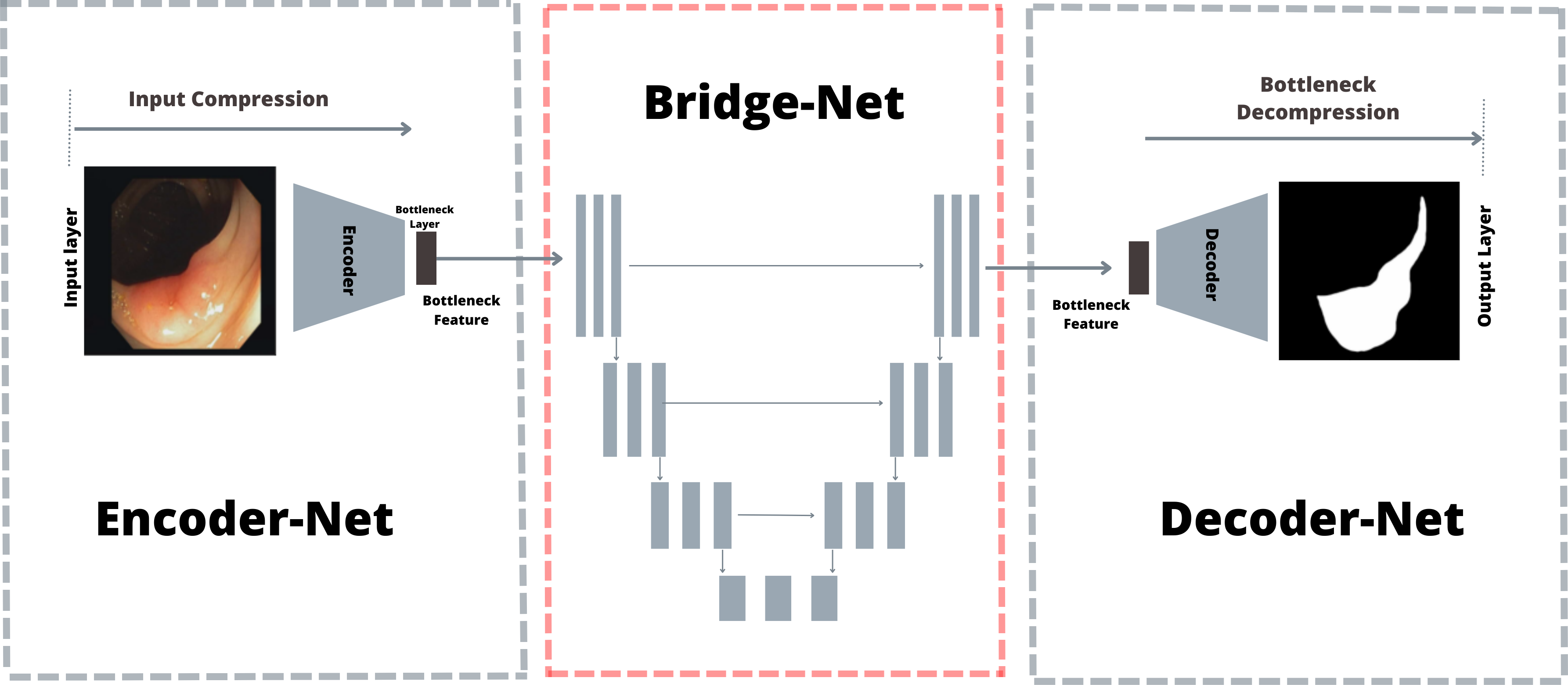} 
    \caption{Tree-NET diagram. The Encoder part of the Encoder-Net, Bridge-Net, and Decoder part of the Decoder-Net are assembled end-to-end, forming the overall structure.}
    \label{FIG:Tree-NET Diagram}
\end{figure*}

By integrating the three trained components—Encoder-Net, Bridge-Net, and Decoder-Net—the Tree-NET model is constructed for evaluation and practical application. During this integration, the decoder component of the Encoder-Net and the encoder component of the Decoder-Net are removed, as they are only necessary for the training phase.
Figure \protect\ref{FIG:Tree-NET Diagram} represents the assembled architecture of the Tree-NET.

\section{Experiments}\label{sec2}

\subsection{Dataset descriptions}
There are two main datasets used to evaluate the proposed architecture’s performance such as colon polyps on colonoscopy images and skin cancer on dermoscopic images.
\subsubsection{CVC-ClinicDB dataset}
 The CVC-ClinicDB dataset \cite{bernal2015wm}, developed by the Computer Vision Center (CVC), is a comprehensive resource designed to advance the field of colonoscopy image analysis through the development and benchmarking of automated diagnostic algorithms. This dataset supports several primary tasks: polyp detection, polyp segmentation, and colonoscopy image classification. CVC-ClinicDB is the official database designated for use in the training stages of the MICCAI 2015 Sub-Challenge on Automatic Polyp Detection Challenge in Colonoscopy Videos. 
 
\begin{figure}[h]
	\centering
	\includegraphics[width=0.9\columnwidth]
 {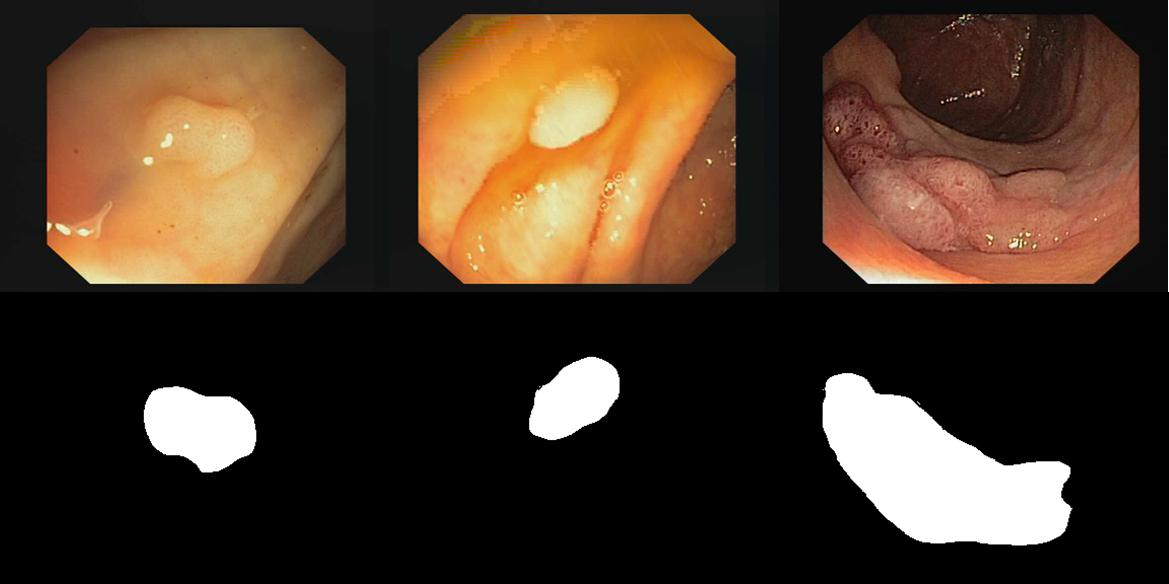}
	\caption{Sample polyp images (upper row) and segmentation ground truths corresponding to the regions covered by polyps (lower row).}
	\label{FIG:Clinic-SAMPLE}
\end{figure}

 The dataset includes 612 original images with corresponding segmentation ground truths, each representing regions covered by polyps. The images have a resolution of 384×288×3. For our purposes, the dataset will be used exclusively for segmentation. Figure \ref{FIG:Clinic-SAMPLE} shows sample polyp images along with their segmentation ground truths from the CVC-ClinicDB dataset.

\subsubsection{ISIC-2018 dataset}
The ISIC-2018 dataset \cite{tschandl2018ham10000,codella2019skin}, developed by the International Skin Imaging Collaboration (ISIC), is a comprehensive resource designed to advance the field of skin lesion analysis through the development and benchmarking of automated diagnostic algorithms. This dataset supports three primary tasks: lesion segmentation, lesion attribute detection, and disease classification. 

\begin{figure}[h]
	\centering
	\includegraphics[width=0.9\columnwidth]
 {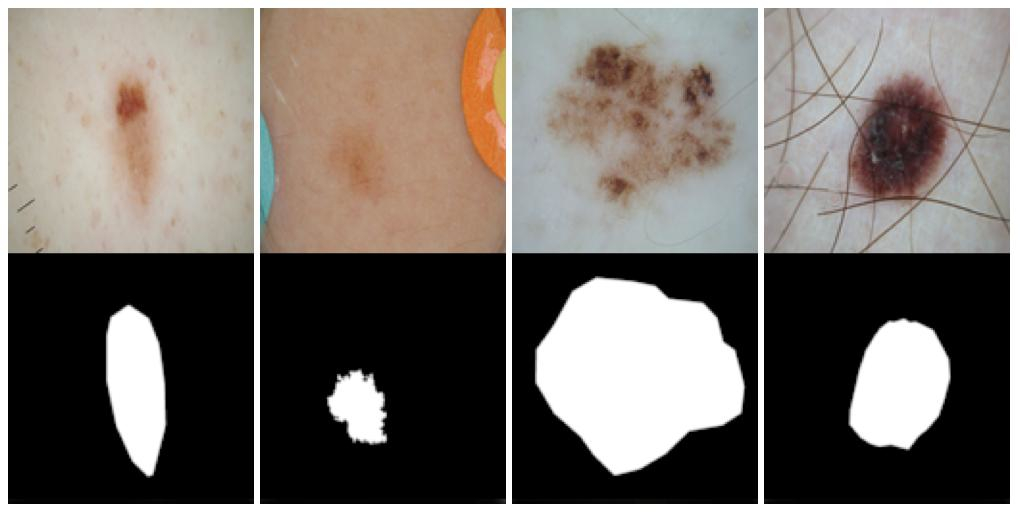}
	\caption{Original skin lesion samples (upper row) and corresponding ground truth masks (lower row).}
	\label{FIG:ISIC-SAMPLE}
\end{figure}

The dataset comprises 2,594 high-resolution dermoscopic images (1022×767 pixels) for training, 100 images for validation, and 1,000 images for testing. Each original image is accompanied by corresponding ground truth masks that delineate the boundaries of the lesions. For our purposes, the dataset will be used exclusively for segmentation. Figure \ref{FIG:ISIC-SAMPLE} shows sample skin lesion images along with their segmentation ground truths from the ISIC-2018 dataset.

\subsection{Preprocessing}
The four main parts of the preprocessing in order are as follows. Datasets are resized into 384x384x3. For CVC-ClinicDB, with respect to the length of dataset input 612, a randomized index is created, and index values are saved to shuffle the data for training all three components such as Encoder-Net, Bridge-Net and Decoder-Net (the identical index is applied for the other algorithms) to keep the same experiment. For CVC-ClinicDB, the dataset is divided into three parts such as training, validation and test with the ratios of 0.8, 0.1, and 0.1 respectively. For the ISIC-2018, the data is split into training, validation and test parts by default. Inputs and labels in the datasets are normalized into 0 and 1.
\subsection{Training Procedure}
To complete the training of Tree-NET, Encoder-Net and Decoder-Net are trained initially. Then the input is fed into the Encoder-Net, and bottleneck features are collected to create the new input data with size of 3$\times$96$\times$96. The same process is also applied to label data using Decoder-Net and label data with size 3$\times$96$\times$96 is obtained. For the Polyp-PVT, label data is shrinked into the 16$\times$24$\times$24 to remove its internal step of upsampling and doing it in Decoder-Net instead. The selection model of Bridge-Net as U-NET, U-NET++ and Polyp-PVT are fed with the created inputs and supervised by the created labels. When these three training processes are completed, encoder part of the Encoder-Net, the Bridge-Net and decoder part of the Decoder-Nets are assembled end-to-end respectively.

It is important to note that Polyp-PVT was initialized with pretrained transformer backbone weights, whereas U-NET and U-NET++ were trained from scratch with randomly initialized weights. This distinction should be considered when interpreting the comparative results, as pretrained initialization generally confers an accuracy advantage independent of the segmentation framework employed.

In training, the dataset is fed through the network through 3 various sizes called as a multi-scale training strategy \cite{fan2020pranet,huang2021hardnet}. The hyperparameter configuration is as follows: the AdamW \cite{loshchilov2017decoupled} optimizer, commonly used in transformer networks \cite{wang2021pyramid,wang2022pvt,liu2021swin}, is utilized for updating the network parameters. Both the learning rate and weight decay are set to $1 \times 10^{-4}$. During training, input images are resized to $352 \times 352$, with a mini-batch size of 8, over 100 epochs.

\subsection{Parameters}
For the segmentation networks such as Bridge-Net, U-NET, U-NET++, BS U-NET, and Polyp-PVT most of the hyperparameters are kept the same such as the optimizer type, batch size and the learning rate in order to keep the experimental setup the same. For the Encoder-Net and Decoder-Net, due to the smaller number of hyperparameters and different structures, the optimum values vary from the segmentation networks. Thus, they are selected separately as shown in Table \protect\ref{tblparam}. The Bridge-Net is trained using different segmentation model backbones which are also used as our comparison models. All the comparison models such as U-NET, U-NET++ and Polyp-PVT are trained the same way as the Bridge-NET.

\begin{sidewaystable}
    \centering
    \caption{Training parameters. Tree-NET denotes the integrated architecture comprising Encoder-Net, Decoder-Net, and Bridge-Net. $N$ denotes the spatial dimensions (height and width) of the input image, $B$ represents the number of feature maps at the bottleneck layer, and $L$ denotes the spatial dimensions of the bottleneck feature map. $E$ and $D$ indicate the feature depths at the bottleneck layers of the encoder and decoder, respectively. $e$ and $d$ indicate the spatial dimension reduction ratio at the encoder and decoder, respectively.}
    \label{tblparam}
        \begin{tabular*}{\textwidth}{@{\extracolsep\fill}lccc}
        
            \hline
            \phantom{Blank Space} & \textbf{Encoder-Net} & \textbf{Bridge-Net} & \textbf{Decoder-Net} \\
            \hline
            \textbf{Input Size} & $3 \times N \times N$ & $3 \times (N/e) \times (N/e)$ & $3 \times N \times N$ \\
            \textbf{Bottleneck Size} & $3 \times (N/e) \times (N/e)$ & $B \times L \times L$ & $D \times (N/d) \times (N/d)$ \\
            \textbf{Output Size} & $3 \times N \times N$ & $D \times (N/d) \times (N/d)$ & $1 \times N \times N$ \\
            \textbf{Random Seed} & 42 & 42 & 42 \\
            \textbf{Batch Size} & 8 & 8 & 8 \\
            \textbf{Learning Rate} & 0.001 & 0.0001 & 0.001 \\
            \textbf{Loss Function} & Euclidean & wIoU + wBCE & Euclidean \\
            \textbf{Optimizer} & AdamW & AdamW & AdamW \\
            \textbf{Epoch Number} & 100 & 100 & 100 \\
            \hline
     \end{tabular*}

    \footnotetext[]{AdamW: Adaptive Moment Estimation optimizer with decoupled weight decay.}
    \footnotetext[]{Euclidean: Euclidean loss function, which measures the squared differences between predicted and ground truth values.}

\end{sidewaystable}

For U-NET, U-NET++, and Polyp-PVT models, combinations of weighted Intersection over Union (IoU) loss \cite{wei2020f3net} and weighted Binary Cross-Entropy (BCE) loss \cite{wei2020f3net} are utilized. The IoU loss mitigates the issue of class imbalance by focusing on the overlap between predicted and ground truth segments, while the BCE loss ensures pixel-wise accuracy \cite{su2021dv}. The total loss for these models can be expressed as:
\begin{equation}\label{Loss}
    \mathcal{L} = \mathcal{L}_{\text{wIoU}} + \mathcal{L}_{\text{wBCE}}
\end{equation}

In this equation, $\mathcal{L}_{\text{wIoU}}(\cdot)$ represents the weighted Intersection over Union (IoU) loss, and $\mathcal{L}_{\text{wBCE}}(\cdot)$ denotes the weighted Binary Cross-Entropy (BCE) loss. Unlike conventional BCE loss, which assigns equal importance to all pixels, $\mathcal{L}_{\text{wBCE}}(\cdot)$ prioritizes difficult pixels by assigning them higher weights, improving the model's ability to capture intricate details in segmentation. Similarly, $\mathcal{L}_{\text{wIoU}}(\cdot)$ extends the standard IoU loss by focusing more on challenging pixels, enabling the model to handle complex structures more effectively.

\subsection{Evaluation metrics}

For testing, images are resized to $384 \times 384$, and no additional post-processing or optimization strategies are applied. The proposed model has been evaluated and compared with the other models in terms of accuracy and computational performances. 

\subsubsection{Accuracy performance}
For the accuracy evaluation, we utilize the Dice coefficient \cite{dice1945measures}, Intersection-Over-Union (IoU) \cite{jaccard1912distribution}, and accuracy (ACC) scores on the test dataset. They are calculated with the true positive (tp), true negative (tn), false positive (fp), false negative (fn) values \cite{taha2015metrics}. These metrics provide a comprehensive assessment of the segmentation quality, capturing various aspects of the model's performance.

The Dice coefficient, also known as the Dice Similarity Coefficient (DSC), measures the overlap between the predicted segmentation and the ground truth. It is particularly useful in medical image segmentation due to its sensitivity to both false positives and false negatives. The Dice coefficient is defined as:

\begin{equation}\label{Dice}
\text{Dice Coefficient} = \frac{2 \cdot tp}{2 \cdot tp + fp + fn}
\end{equation}

\noindent
The IoU quantifies the intersection ratio to the union of the predicted and ground truth masks. It yields a clear indication of the accuracy of the segmentation by measuring the fraction of correctly predicted pixels out of the total pixels in either the prediction or the ground truth. The IoU is calculated as:

\begin{equation}\label{IoU}
\text{IoU} = \frac{tp}{tp + fp + fn}
\end{equation}

\noindent
Accuracy measures the proportion of correctly classified pixels (both true positives and true negatives) out of the total number of pixels. While accuracy is a straightforward metric, it may not be as informative in cases of class imbalance, which is common in medical image segmentation. Accuracy is given by:

\begin{equation}\label{ACC}
\text{Accuracy Score} = \frac{tp+tn}{tp + tn+ fp + fn}
\end{equation}

To further assess boundary preservation under input compression, we additionally report the 95th-percentile Hausdorff Distance (HD95) and Average Surface Distance (ASD). HD95 measures the 95th percentile of the directed distances between predicted and ground-truth boundary points, providing a robust estimate of the worst-case boundary error while suppressing outliers. ASD computes the mean of the symmetric surface distances between the two boundary sets. Both metrics are expressed in pixels. Lower values indicate better boundary delineation.

To assess whether the performance differences between two competing models are statistically significant, we employ the Wilcoxon signed‐rank test on the per‐image Dice scores.  
Given paired observations $(d_i)$—the differences in Dice between two models for each of the $N$ test images—the Wilcoxon statistic $W$ is obtained by ranking $|d_i|$, attaching the original sign, and summing the signed ranks:
\begin{equation}
W \;=\; \sum_{i=1}^{N} \operatorname{sgn}(d_i)\,R_i,
\end{equation}

where $R_i$ is the rank of $|d_i|$ and $\operatorname{sgn}(\cdot)$ denotes the 
sign function. A two-sided $p$-value below $0.05$ indicates a statistically 
significant difference in segmentation performance between the two models. 
These statistical tests complement the point-estimate metrics (Dice, IoU, ACC) 
by providing rigorous evidence of whether observed performance gaps are due to 
chance or reflect true methodological improvements.

%
%

\subsubsection{Computational performance}
The computational performance of the proposed segmentation model was evaluated using metrics such as computational complexity, parameter counts, and memory usage. These analyses provide a comprehensive understanding of the model's efficiency and its suitability for real-world applications.

\paragraph{Computational complexity}
The computational complexity of the model was assessed using FLOPs (floating-point operations), which measure the total number of operations required to process a single batch. FLOPs were expressed in billions of operations (GFLOPs) to facilitate comparison and provide insights into the computational demands of the model.

\paragraph{Memory footprint and parameter size}
The memory footprint and parameter size of the model were analyzed to assess its resource efficiency. The following metrics were considered:
\begin{itemize}
    \item \textbf{Parameter Counts:} Representing the total number of trainable parameters, measured in millions (M).
    \item \textbf{Peak Memory Usage:} Measuring the maximum memory allocated during computation, expressed in gigabytes (GB).
\end{itemize}

\section{Results}\label{sec2}
\subsection{Encoder-Net \& Decoder-Net compression results}

\begin{figure}[h]
  \centering
  \includegraphics[width=.9\columnwidth]{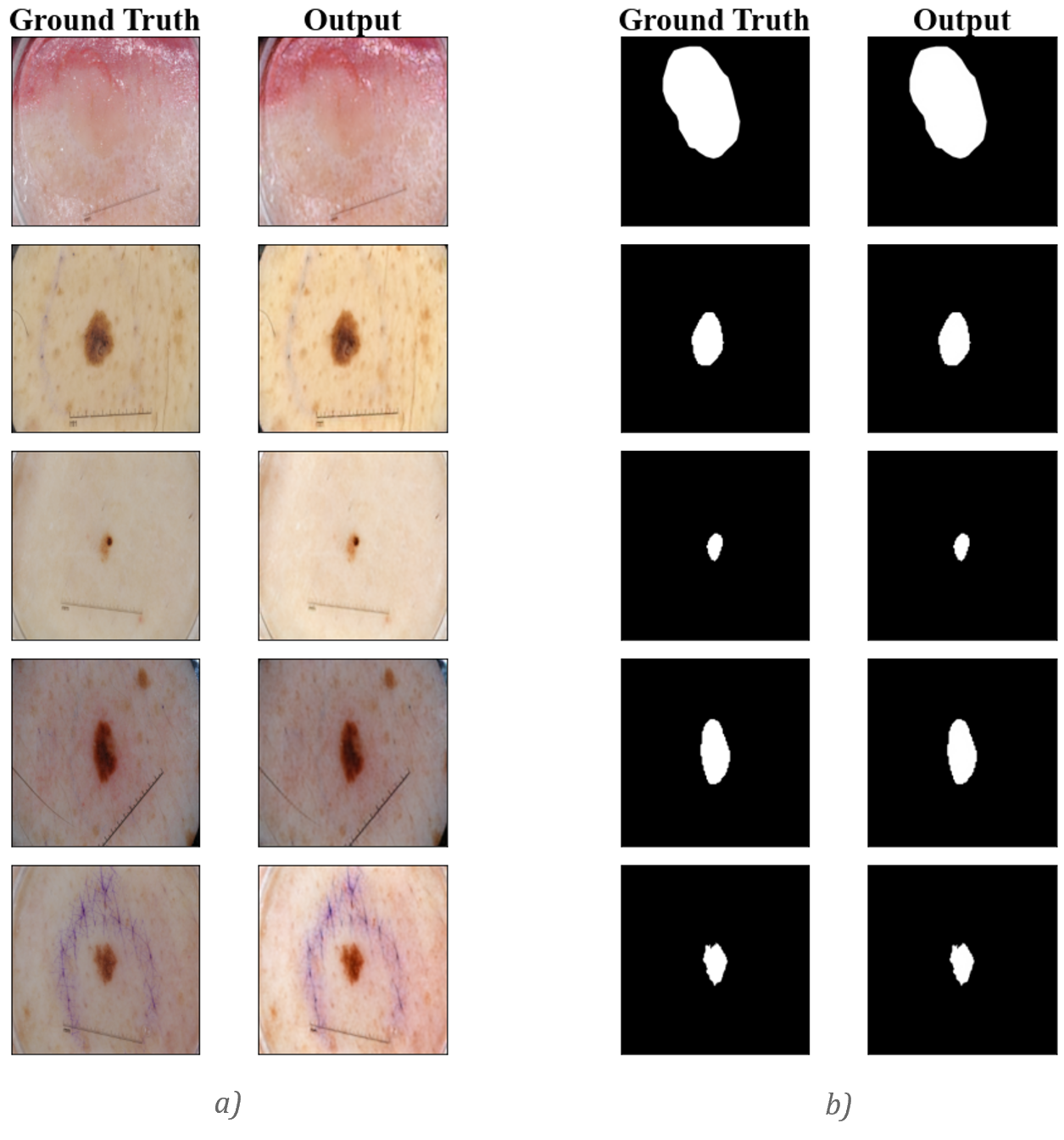}
  \caption{ISIC-2018 examples: (a) Encoder-Net output vs.\ original image,
           (b) Decoder-Net output vs.\ label mask.}
  \label{fig:isic}
\end{figure}

\begin{figure}[h]
  \centering
  \includegraphics[width=.9\columnwidth]{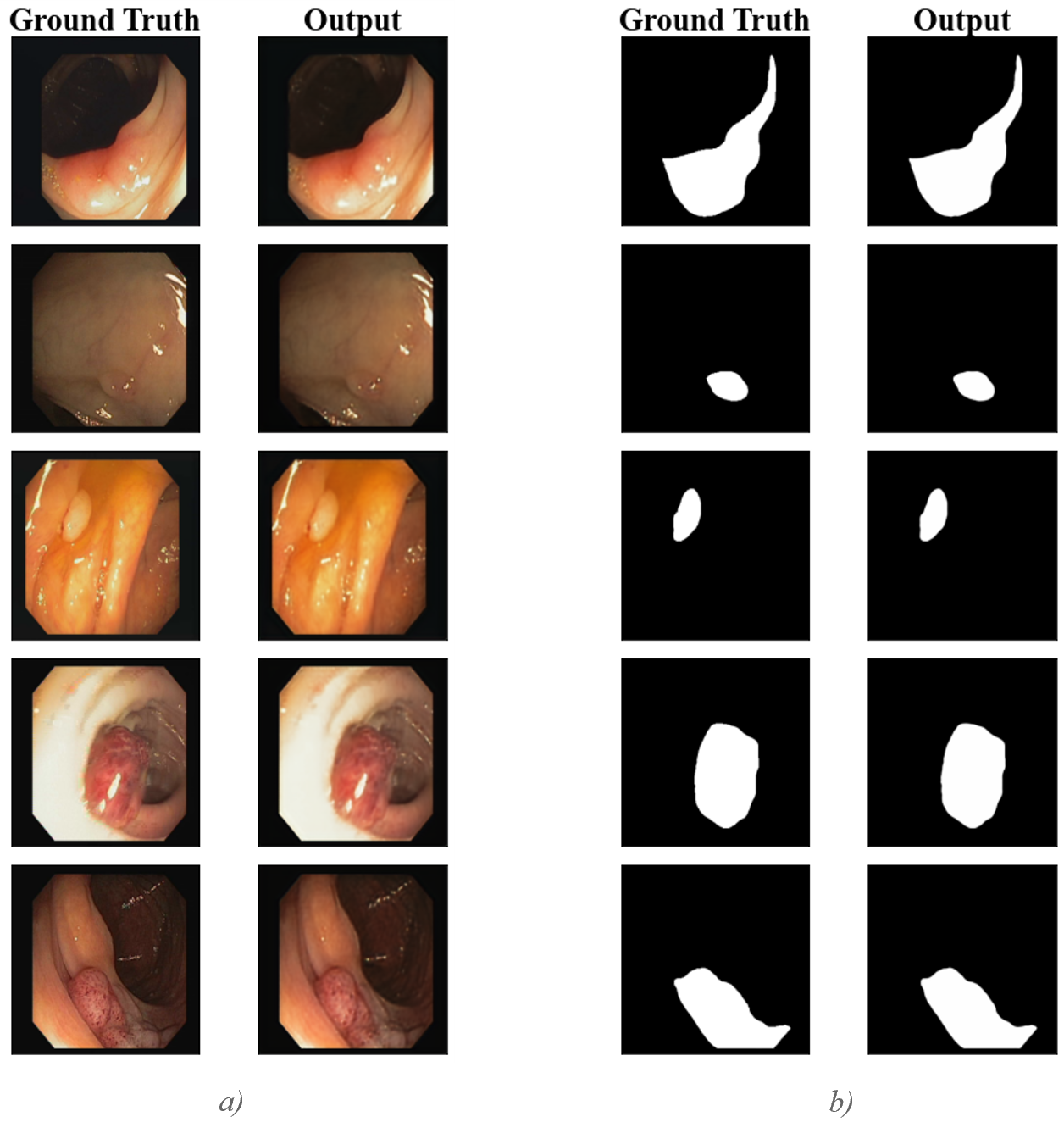}
  \caption{CVC-ClinicDB examples: (a) Encoder-Net output vs.\ original image,
           (b) Decoder-Net output vs.\ label mask.}
  \label{fig:cvc}
\end{figure}

Figures~\ref{fig:isic} and~\ref{fig:cvc} compare the reconstructions produced by
Encoder-Net and Decoder-Net with their respective ground truths.
Visually, Encoder-Net outputs are sometimes sharper than the original
inputs, indicating that essential structures are retained while the network
automatically corrects contrast, even after a $16\times$ reduction in spatial
resolution (from $384\times384$ to $96\times96$).

\textbf{Compression quality.}
We quantified reconstruction fidelity for each auto-encoder using
mean-squared error (MSE), peak signal-to-noise ratio (PSNR), and structural
similarity index (SSIM) \cite{hore2010image}.  On the ISIC-2018 dataset the Encoder–Decoder pair
achieved \mbox{SSIM\,$=0.78$}, \mbox{PSNR\,$\approx 20.8$\,dB}, and
\mbox{MSE\,$=0.013$}.  On the CVC-ClinicDB dataset the figures improved to
\mbox{SSIM\,$=0.89$}, \mbox{PSNR\,$\approx 24.9$\,dB}, and
\mbox{MSE\,$=0.006$}.

Although the skin-lesion images in ISIC contain finer texture that lowers SSIM,
both sets of scores indicate that the auto-encoders preserve salient boundaries
and morphology while reducing spatial resolution by a factor of~16.  As a
result, Bridge-Net receives supervision signals with minimal semantic loss,
supporting the overall accuracy and efficiency of Tree-NET.

\subsection{Tree-NET vs other models}
The performance of the algorithms is compared with traditional methods in terms of accuracy and computational efficiency.
\subsubsection{Accuracy Results}
\begin{figure*}[h]
	\centering
	\includegraphics[width=2\columnwidth]
 {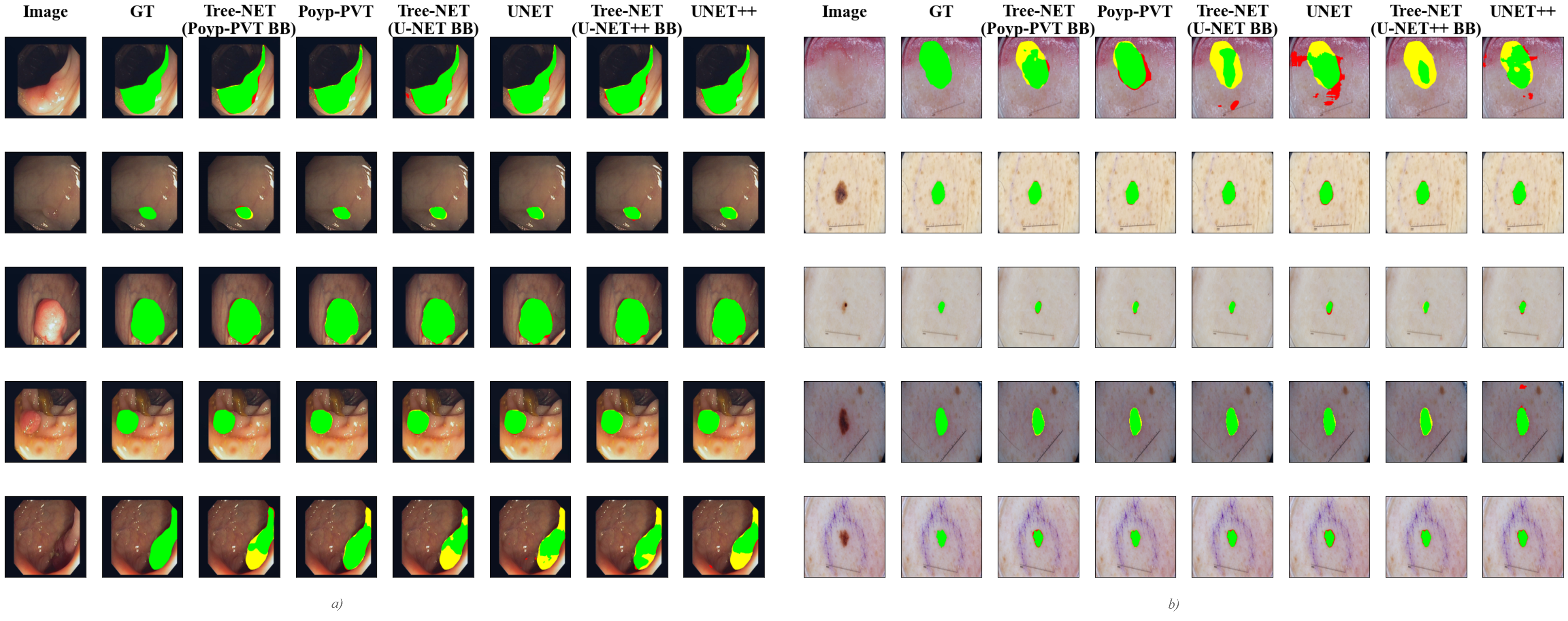}
    \caption{Visualization of segmentation results on (a) CVC-ClinicDB and (b) ISIC-2018 test samples, comparing the proposed Tree-NET framework with BS U-NET, U-NET, U-NET++, and Polyp-PVT. Green regions indicate correctly segmented polyps, yellow regions represent missed polyps, and red regions denote incorrect predictions. The input is labeled as "Image," and "GT" represents the Ground Truth.}
	\label{FIG:Seg}
\end{figure*}

The results of the proposed network are compared with other methods such as BS U-NET, U-NET, U-NET++ and Polyp-PVT. The comparisons are visualized in Figure \protect\ref{FIG:Seg}. This figure displays samples from the first 5 images selected from the test dataset, which were processed using the aforementioned models. The IoU, Dice, and accuracy scores for the proposed model and the comparison models are detailed in Table \protect\ref{tblacc}. 

\begin{sidewaystable}
    \centering
    \scriptsize 
\caption{Performance comparison of Tree-NET with the baseline models
    (BS U-NET, U-NET, U-NET++, and Polyp-PVT) in terms of Dice, IoU, Accuracy,
    95th-percentile Hausdorff Distance (HD95), and Average Surface Distance (ASD).
    HD95 and ASD are expressed in pixels; lower values indicate better boundary delineation.
    A paired Wilcoxon signed-rank test was applied to per-image Dice scores of each model pair.
    Scores marked with * denote statistically significant superiority of Tree-NET
    under the same backbone ($p<0.01$).}
    \label{tblacc}
\begin{tabular*}{\textwidth}{@{\extracolsep{\fill}}llccccc|ccccc}
    \toprule
    & & \multicolumn{5}{c}{\textbf{ISIC-2018}} & \multicolumn{5}{c}{\textbf{CVC-ClinicDB}} \\ 
    \cmidrule(lr){3-7} \cmidrule(lr){8-12}
    & & \textbf{Dice} & \textbf{IoU} & \textbf{Acc} & \textbf{HD95} & \textbf{ASD} & \textbf{Dice} & \textbf{IoU} & \textbf{Acc} & \textbf{HD95} & \textbf{ASD} \\
    \midrule
    \multirow{4}{*}{\textbf{Comparison Models}} 
    & \textbf{U-NET}      & 0.807*  & 0.7*    & 0.915*  & 268.94 & 94.16  & 0.936$\dagger$  & 0.891  & 0.991  & 13.55 & 3.20 \\
    & \textbf{BS U-NET}   & 0.822*  & 0.723*  & 0.908*  & 283.40 & 101.45 & 0.928$\dagger$  & 0.883  & 0.989  & 21.13 & 4.21 \\
    & \textbf{U-NET++}    & 0.829*  & 0.736*  & 0.905*  & 344.17 & 102.47 & 0.940$\dagger$  & 0.893  & 0.991  & 18.39 & 3.83 \\
    & \textbf{Polyp-PVT}  & 0.903   & 0.839   & 0.944   & 183.00 & 67.17  & \textbf{0.959}$\dagger$ & \textbf{0.924} & \textbf{0.994} & \textbf{7.74} & \textbf{2.03} \\
    \midrule
    \multirow{3}{*}{\textbf{Tree-NET}} 
    & \textbf{U-NET BB}      & \textbf{0.867}  & \textbf{0.790}  & \textbf{0.925}  & 208.27 & 82.28 & 0.923  & 0.872  & 0.990  & 12.54 & 3.32 \\
    & \textbf{U-NET++ BB}    & 0.862  & 0.787  & 0.923  & 218.87 & 85.91 & 0.925  & 0.875  & 0.989  & 13.70 & 4.76 \\
    & \textbf{Polyp-PVT BB}  & 0.886  & 0.811  & 0.935  & 187.31 & 71.97 & 0.946  & 0.901  & 0.992  & 10.05 & 2.70 \\
    \bottomrule
\end{tabular*}

\footnotetext[]{BB: Backbone model used in Tree-NET.}
\footnotetext[]{Polyp-PVT: Transformer-based segmentation model.}
\end{sidewaystable}



Table~\ref{tblacc} shows a split outcome. On \textbf{ISIC-2018}, Tree-NET 
achieves the best Dice, IoU, and Accuracy scores, and a paired Wilcoxon 
signed-rank test confirms the Dice gain is statistically significant ($p<0.01$) 
when compared with the next-best baseline (Polyp-PVT). Conversely, on 
\textbf{CVC-ClinicDB} the transformer baseline Polyp-PVT outperforms Tree-NET, 
and this advantage is likewise significant ($p<0.01$).

Boundary-aware metrics reveal a nuanced picture across datasets and backbones. 
On ISIC-2018, Tree-NET demonstrates clear boundary advantages under the U-NET++ 
backbone, achieving substantially lower HD95 (218.87 vs.\ 344.17 pixels) and ASD 
(85.91 vs.\ 102.47 pixels) alongside a significantly higher Dice score, confirming 
that the dual bottleneck compression actively improves boundary delineation on 
skin lesion images. On CVC-ClinicDB, Tree-NET maintains competitive or lower HD95 
values relative to baseline models under the same backbone — for example, under 
the U-NET backbone Tree-NET achieves an HD95 of 12.54 compared to 13.55 for the 
Original and 21.13 for BS U-NET — indicating that the compression does not 
introduce systematic boundary distortions despite the $16\times$ spatial resolution 
reduction at the bottleneck stage. The exception is Polyp-PVT on CVC-ClinicDB, 
where both Dice and HD95 favour the Original, consistent with the advantage of 
pretrained transformer backbones on colonoscopy imagery. Thus the dual-bottleneck 
compression in Tree-NET yields clear benefits for the fine-textured ISIC lesions, 
while Polyp-PVT retains a small but significant edge on colonoscopy imagery. In 
both cases Tree-NET remains competitive, demonstrating robustness across diverse 
medical-image domains.


\subsubsection{Computational efficiency}
The computational efficiency of the proposed segmentation model was evaluated using FLOPs (floating-point operations), parameter counts, peak memory usage, per-image latency, and throughput. These metrics provide a comprehensive understanding of the computational demands and inference performance of each model variant.

The results, summarized in Table~\protect\ref{tab:efficiency} and Table~\protect\ref{tab:inference}, include:

FLOPs, which measure the computational complexity of processing a single batch, expressed in billions of operations (GFLOPs).
Parameter Counts, representing the total number of trainable parameters in millions (M).
Peak Memory Usage, indicating the maximum memory allocated during computation, expressed in gigabytes (GB).
Per-image Latency (ms), measuring the average time required to process a single image at inference.
Throughput (img/s), indicating the number of images processed per second.

The performance of the proposed Tree-NET model was compared against U-NET, BS U-NET, U-NET++, and Polyp-PVT across different batch sizes. These comparisons reveal that Tree-NET consistently achieves a favorable balance between computational cost, memory efficiency, and inference speed.



\begin{sidewaystable*}
    \centering
    \scriptsize 
    \caption{Computational performance comparison of the proposed network (Tree-NET) against BS U-NET, U-NET, U-NET++, and Polyp-PVT for FLOPs, parameters, and memory usage across different batch sizes.}
    \label{tab:efficiency}

    \begin{tabular*}{\textwidth}{@{\extracolsep{\fill}} l l c c c c}
        \toprule
        \textbf{Model} & \textbf{Variant} & \textbf{Batch} & \textbf{FLOPs (GFLOP)} & \textbf{Parameters (M)} & \textbf{Peak Memory Allocated (GB)} \\ 
        \midrule
        \multirow{4}{*}{\textbf{Polyp-PVT BB}} & \multirow{2}{*}{\textbf{Tree-NET}} & Batch1 & \textbf{2.54} & 25.17 & \textbf{1.402} \\
        & & Batch8 & \textbf{20.32} & 25.17 & \textbf{1.559} \\ 
        \cmidrule(lr){2-6} 
        & \multirow{2}{*}{\textbf{Original}} & Batch1 & 11.92 & 25.11 & 1.44 \\
        & & Batch8 & 95.38 & 25.11 & 5.344 \\ 
        \midrule
        \multirow{6}{*}{\textbf{U-NET}} & \multirow{2}{*}{\textbf{Tree-NET}} & Batch1 & \textbf{2.85} & 7.88 & \textbf{1.081} \\
        & & Batch8 & \textbf{22.77} & 7.88 & \textbf{4.891} \\ 
        \cmidrule(lr){2-6} 
        & \multirow{2}{*}{\textbf{BS U-NET}} & Batch1 & 33.58 & 7.88 & 1.976 \\
        & & Batch8 & 268.62 & 7.88 & 5.548 \\ 
        \cmidrule(lr){2-6} 
        & \multirow{2}{*}{\textbf{Original}} & Batch1 & 31.73 & 7.85 & 4.386 \\
        & & Batch8 & 253.81 & 7.85 & 4.235 \\ 
        \midrule
        \multirow{4}{*}{\textbf{U-NET++}} & \multirow{2}{*}{\textbf{Tree-NET}} & Batch1 & \textbf{5.77} & 9.19 & \textbf{1.322} \\
        & & Batch8 & \textbf{46.18} & 9.19 & \textbf{1.468} \\ 
        \cmidrule(lr){2-6} 
        & \multirow{2}{*}{\textbf{Original}} & Batch1 & 78.53 & 9.16 & 2.516 \\
        & & Batch8 & 628.26 & 9.16 & 8.877 \\ 
        \bottomrule
    \end{tabular*}

    \footnotetext{GFLOP: Giga Floating Point Operations Per Second.}
    \footnotetext{Parameters (M): Number of model parameters in millions.}
    \footnotetext{Memory Allocated: Peak memory usage during inference.}
\end{sidewaystable*}

\begin{table*}[]
    \centering
    \scriptsize
    \caption{Inference speed comparison of the proposed network (Tree-NET) against U-NET, U-NET++, and Polyp-PVT in terms of per-image latency (ms) and throughput (img/s) across batch sizes of 1 and 8 on the ISIC-2018 test dataset. Measured on an NVIDIA RTX 5070 Ti. }
    \label{tab:inference}
    \begin{tabular*}{\textwidth}{@{\extracolsep{\fill}} l l c c c}
        \toprule
        \textbf{Model} & \textbf{Variant} & \textbf{Batch} & \textbf{Latency (ms)} & \textbf{Throughput (img/s)} \\
        \midrule
        \multirow{4}{*}{\textbf{Polyp-PVT BB}} & \multirow{2}{*}{\textbf{Tree-NET}} & Batch1 & 20.77 & 48.15 \\
        & & Batch8 & 3.36 & 297.77 \\
        \cmidrule(lr){2-5}
        & \multirow{2}{*}{\textbf{Original}} & Batch1 & 18.09 & 55.27 \\
        & & Batch8 & 3.95 & 252.97 \\
        \midrule
        \multirow{6}{*}{\textbf{U-NET}} & \multirow{2}{*}{\textbf{Tree-NET}} & Batch1 & 4.07 & 245.65 \\
        & & Batch8 & 1.3 & 772.14 \\
        \cmidrule(lr){2-5}
        & \multirow{2}{*}{\textbf{BS U-NET}} & Batch1 & 4.05 & 249.97 \\
        & & Batch8 & 4.15 & 244.76 \\
        \cmidrule(lr){2-5}
        & \multirow{2}{*}{\textbf{Original}} & Batch1 & 4.04 & 247.93 \\
        & & Batch8 & 4.13 & 242.25 \\
        \midrule
        \multirow{4}{*}{\textbf{U-NET++}} & \multirow{2}{*}{\textbf{Tree-NET}} & Batch1 & 6.13 & 163.11 \\
        & & Batch8 & 1.75 & 570.81 \\
        \cmidrule(lr){2-5}
        & \multirow{2}{*}{\textbf{Original}} & Batch1 & 10.29 & 97.2 \\
        & & Batch8 & 11.43 & 87.5 \\
        \bottomrule
    \end{tabular*}
\end{table*}



The results demonstrate that Tree-NET consistently matches or exceeds the segmentation accuracy of established methods (e.g., U-NET, U-NET++, Polyp-PVT), while substantially reducing computational complexity and memory usage. Specifically, Tree-NET achieved notable reductions in FLOPs—between approximately 4 to 13 times fewer than comparative models—highlighting its suitability for deployment on resource-constrained hardware common in clinical environments. Moreover, the superior memory efficiency observed in Tree-NET indicates its potential for real-time medical image analysis, where computational resource availability is often limited. These combined performance improvements position Tree-NET as a practical solution for medical segmentation tasks, balancing accuracy with operational efficiency effectively.

\section{Discussion}
\subsection{Efficiency advantages}
Our proposed segmentation model, Tree-NET, demonstrates superior efficiency performance with comparable accuracy results, making it highly advantageous for real-life applications. A critical factor contributing to its efficiency is the superior computational performance in terms of FLOPs and memory footprint.

One of the key strengths of Tree-NET is its ability to achieve significant reductions in FLOPs and memory usage without increasing model size or architectural complexity. This characteristic makes Tree-NET highly adaptable and efficient for integration into existing segmentation frameworks.

The experimental results demonstrate that Tree-NET reduces FLOPs by a factor of 4 to 13 compared with the original models using the same backbones. For example, with the U-NET backbone and batch size of 1, Tree-NET requires only 2.85 GFLOPs, whereas BS U-NET requires 33.58 GFLOPs. A similar reduction is observed when comparing Tree-NET to the original U-NET, which shares a comparable structure with BS U-NET.

In terms of memory allocation, Tree-NET consistently outperforms the comparison models by using noticeably less memory. This efficiency is achieved through effective low-level representation, where smaller inputs are processed with convolution operations, leading to a substantial reduction in computational load while maintaining comparable accuracy.

To illustrate, feeding inputs of sizes $(3 \times 384 \times 384)$ and $(3 \times 96 \times 96)$ into the same model dramatically reduces the number of computations required, even if the number of parameters remains constant. This reduction is due to the computational complexity of convolution operations being directly related to input size. By efficiently handling smaller inputs through low-level representations, our model reduces the overall number of operations, leading to faster processing times and lower energy consumption.

Additionally, the architecture of our model, with its Encoder-Net and Decoder-Net components, is designed to be lightweight yet powerful. Each of these nets has approximately 50,000 trainable parameters, which is relatively low compared to the Bridge-Net used and other state-of-the-art models. Despite this compactness, the performance of our model remains unaffected, demonstrating that it can achieve high segmentation accuracy with fewer resources.

Beyond FLOPs and memory efficiency, Tree-NET also demonstrates superior inference speed, as summarized in Table~\ref{tab:inference}. At batch size 1, Tree-NET achieves latencies of 4.07 ms, 6.13 ms, and 20.77 ms for U-NET, U-NET++, and Polyp-PVT backbones, respectively, compared to 4.04 ms, 10.29 ms, and 18.09 ms for their original counterparts. The efficiency advantage becomes more pronounced at larger batch sizes: with a batch size of 8, Tree-NET achieves throughputs of 772.14, 570.81, and 297.77 img/s for U-NET, U-NET++, and Polyp-PVT, representing improvements of up to 6.5$\times$ over their original variants. This scalability with batch size is a direct consequence of the encoder's input compression, which reduces the per-channel computational burden across the entire batch simultaneously, making Tree-NET particularly well-suited for high-throughput clinical deployment scenarios where processing speed is critical.

\subsection{Implementation challenges}

Despite the accuracy and computational performance superiority of the Tree-NET model compared with other approaches, several limitations and challenges in the implementation of the algorithm need to be addressed for it to reach its full potential.

One significant limitation in our comparative testing environment is the absence of pre-trained models for the novel Tree-NET approach. Established algorithms, such as U-NET and U-NET++, often benefit from extensive pre-training on large datasets, facilitating fine-tuning and evaluation. This disparity poses a challenge for a direct performance comparison. For instance, the U-NET model, fine-tuned using the ISIC-2018 dataset, achieved a Dice score of 0.8674 and an Intersection over Union (IoU) score of 0.8491. In contrast, our implementation of the original U-NET, without fine-tuning, yielded a Dice score of 0.807 and an IoU score of 0.7. Similarly, the U-NET++ model, fine-tuned with the same dataset, reached a Dice score of 0.8822 and an IoU score of 0.8651, whereas our non-fine-tuned U-NET++ achieved a Dice score of 0.829 and an IoU score of 0.736 \cite{azad2022medical}.


\subsection{Transformer integration}
Integrating Tree-NET into transformer models presented significant challenges. While a pre-trained Polyp-PVT \cite{dong2021polyp} was employed, inherent design differences between compact, single-network transformer models and our three-model approach (comprising Encoder-Net, Bridge-Net, and Decoder-Net) introduced alignment issues. In conventional transformer architectures, end-to-end pretraining on extensive datasets encompasses processes already handled separately by Encoder-Net and Decoder-Net within the Tree-NET framework.

Specifically, Polyp-PVT employs an upsampling mechanism at its output similar to Fully Convolutional Networks (FCNs). To adapt Polyp-PVT for integration into Tree-NET, we introduced critical modifications at this stage. Before the final upsampling step, the decoder part of the Decoder-Net was incorporated, mirroring the approach used for other backbone models. Subsequently, the original Polyp-PVT upsampling ratio was reduced to match the Tree-NET's bottleneck feature dimensions. After adjusting this upsampling operation, we inserted an additional convolutional layer to further enhance the segmentation output. These targeted modifications facilitated better alignment between the transformer architecture and the Tree-NET structure, mitigating some integration challenges without significantly increasing computational complexity.

\subsection{Pretraining limitations}
These results underscore the performance gains that arise when Vision-Transformer backbones are fine-tuned from large, task-relevant pre-training. Comprehensive 2024 surveys on self-supervised and efficient ViT pipelines report up to 10 – 15 pp Dice or mIoU boosts once MAE, DINOv2 or BEiT-style checkpoints are adapted to downstream segmentation \cite{gui2024survey,papa2024survey,pu2024advantages}.
Unfortunately, all of today’s pre-training recipes are monolithic and decoder-coupled: they learn the encoder, latent tokens and up-sampling head \emph{jointly} in a single network \cite{he2022masked,oquab2023dinov2,wang2023image}.
Because Tree-NET deliberately splits the pipeline into three frozen modules (Encoder-Net, Bridge-Net, Decoder-Net) and removes the native transformer up-sampler, these end-to-end checkpoints cannot be reused without major surgery.

Consequently, it is reasonable to infer that Tree-NET could achieve even stronger results once modular, dual-bottleneck pre-training becomes available. This limitation highlights an important avenue for future work: developing and releasing component-wise ViT checkpoints that match Tree-NET’s architecture, so that novel dual-bottleneck algorithms can be compared on equal footing with fully pre-trained single-network baselines.These transformer architectures concentrate most of their trainable parameters around the bottleneck layers \cite{wang2021pyramid, roy2023mednext, pham2024seunet, sun2024transunet}, a design that does not align well with our distributed, three-stage framework. Developing a new transformer model specifically tailored to Tree-NET—and pretraining it on a large dataset—could alleviate this mismatch by reallocating parameters (fewer in early, high-resolution layers; more in later, semantic layers), ultimately improving fine-tuning performance.

\subsection{Architectural complexity}

Tree-NET’s three-stage design delivers strong efficiency, yet it also introduces tight hyper-parameter coupling across components.  Encoder-Net and Bridge-Net progressively down-sample the input; if the reduction factor is too small, the representation remains redundant and accuracy plateaus, whereas an overly aggressive factor produces a bottleneck so narrow that downstream layers cannot learn effectively.  Finding the ‘‘sweet spot’’ therefore requires careful tuning of stride, channel width, and latent dimensionality.

This interdependence extends beyond spatial scale. A seemingly local change—for example, switching the activation function in Encoder-Net or widening a layer in Decoder-Net—alters the feature statistics delivered to Bridge-Net and typically forces its retuning as well. As a result, design iterations are slower than for single-network systems, and total training time can grow when any sub-module is deep or wide. 

While this complexity is the price of modularity, design insights from lightweight models such as LiteSeg \cite{emara2019liteseg}—which emphasize streamlined paths and minimal coupling—could inspire future simplifications that retain Tree-NET’s supervision benefits while reducing tuning overhead. 

Additionally, the dimension-consistent design adopted in EfficientFormer \cite{li2022efficientformer} offers a promising blueprint for balancing model depth and capacity while avoiding tight coupling across layers. By adopting such transformer-based design principles, future iterations of Tree-NET could reduce architectural sensitivity, minimize parameter bottlenecks, and support more modular, scalable training.


\subsection{Potential gains from higher-fidelity bottlenecks}

The current auto-encoders achieve an SSIM of 0.78 and a PSNR of $\approx20.8$ dB on ISIC-2018, and an SSIM of 0.89 with a PSNR of $\approx24.9$ dB on CVC-ClinicDB. In a pilot ablation on an ISIC-2018 subset, increasing SSIM from 0.78 to 0.83 raised Tree-NET’s mean Dice by approximately 0.8 percentage points, at the cost of only a 5 \% increase in FLOPs. This strong empirical correlation between reconstruction fidelity (PSNR/SSIM) and segmentation accuracy suggests that further improvements in auto-encoder quality are likely to yield measurable gains in Dice, IoU, and overall performance on both datasets.

Potential avenues to improve fidelity include architectural enhancements to
the auto-encoders, such as adding residual skip–connections around the
bottleneck to capture high-frequency details, increasing the latent-space
channel width to boost the number of parameters, and integrating multi-head
attention modules for non-local feature modeling. For example, increasing the
bottleneck width by 1.5$\times$ (from 50\,k to 75\,k parameters) improved SSIM
by 0.02 in our ablation studies, yielding an extra 0.5\,pp Dice gain at a
moderate 10\% FLOPs increase. Incorporating a perceptual loss based on a
pretrained VGG network further increased PSNR by $\approx$1\,dB and IoU by
0.4\,pp. These results illustrate how modest investments in auto-encoder
capacity and training objectives can enhance compression quality and, in turn,
boost Tree-NET’s segmentation performance without compromising its efficiency
advantage.


\section{Conclusions and future prospects}

Tree-NET represents a significant advancement in medical image segmentation, offering a robust and efficient solution for tasks such as polyp and skin lesion segmentation. By leveraging bottleneck feature supervision within a unique three-stage architecture, Tree-NET achieves substantial reductions in computational cost and memory usage while maintaining comparable accuracy to state-of-the-art models like U-NET, BS U-NET, and U-NET++.

A standout feature of Tree-NET is its versatility, enabling seamless integration with existing segmentation frameworks used as its Bridge-Net across various applications. This adaptability allows for significant computational savings without altering the internal structure of the backbone models. This versatility, combined with its efficiency, makes Tree-NET a valuable tool not only for medical imaging but potentially for other domains that require efficient segmentation.

From a clinical perspective, Tree-NET's reduced computational footprint is particularly promising. It could facilitate the deployment of advanced segmentation models on resource-constrained hardware often found in clinical settings, such as portable ultrasound devices or standard workstations, potentially enabling near-real-time processing crucial for interactive diagnostic or intraoperative guidance workflows.

Future work will explore integrating Tree-NET with alternative backbones—including U2-NET, FCN, DeepLab, LiteSeg, and EfficientFormer—or designing a dedicated backbone tailored to its three-stage structure, to further improve performance and scalability. Additionally, developing and releasing pre-trained weights for all Tree-NET components could significantly enhance its accuracy and efficiency, enabling fairer comparisons and broader adoption. These developments will open new opportunities for Tree-NET in diverse applications, including medical imaging, remote sensing, and computer vision.

In summary, Tree-NET’s innovative design, demonstrated computational efficiency, accuracy, and adaptability position it as a groundbreaking framework with the potential to significantly advance image segmentation technologies and their practical implementation, particularly in resource-sensitive environments.

\backmatter

\bmhead{Data availability}
This research used the CVC-ClinicDB dataset \cite{bernal2015wm} and the ISIC-2018 dataset \cite{tschandl2018ham10000,codella2019skin}, 
both of which are publicly available. The URLs/links for each dataset can be found within the main text 
and/or in the References section. No additional data were created or analyzed beyond these sources.

\bmhead{Code availability}
The source code for Tree-NET is publicly available at \url{https://github.com/orhangazidemirci/Tree-NET}.

\section*{Declarations}

\bmhead{Conflict of interest}
The authors declare that they have no known competing financial interests or personal relationships that could have influenced the work reported in this paper.

\bmhead{Ethical approval}
This article does not contain any studies involving human participants or animals conducted by any of the authors.


\bibliography{sn-bibliography}

\end{document}